\documentclass[letterpaper,prl,floatfix,twocolumn,aps,10pt]{revtex4-1}

\usepackage{
	amssymb, 
	amsthm, 
	graphicx, 
	xspace,
	overpic,
	ifthen,
	verbatim, 
	subfigure, 
	appendix,
	ifthen,
	nicefrac, 
	color, 
	tensor, 
	etoolbox, 
	centernot, 
	MnSymbol, 
	amsmath,
	bm
} 

\newcommand{\ket}[1]{\lvert #1 \rangle  }

\newcommand{\braket}[2]{  \langle #1 \vert #2 \rangle  }

\newcommand{\abs}[1]{| #1 |} 

\newcommand{\dd}{\mathrm{d}} 

\newcommand{\refstyle}{2}  
\newcommand{\eqrefstyle}{2}  
\newcommand{\Jeqref}[1] {\ifnum\refstyle=1{Eq.~(\ref{#1})}\else{equation (\ref{#1})}\fi}
\newcommand{\Jeqsref}[1] {\ifnum\refstyle=1{Eqs.~(\ref{#1})}\else{equations (\ref{#1})}\fi}
\newcommand{\EQref}[1] {\ifnum\eqrefstyle=1{(\ref{#1})}\else{\Jeqref{#1}}\fi}
\newcommand{\EQsref}[1] {\ifnum\eqrefstyle=1{(\ref{#1})}\else{\Jeqsref{#1}}\fi}
\newcommand{\Eqref}[1]  {\ifnum\eqrefstyle=3{\Jeqref{#1}}\else{(\ref{#1})}\fi}
\newcommand{\Eqsref}[1]  {\ifnum\eqrefstyle=3{\Jeqsref{#1}}\else{(\ref{#1})}\fi}
\newcommand{\Figref}[1]{\ifnum\refstyle=1{Fig.~\ref{#1}}\else{figure~\ref{#1}}\fi}
\newcommand{\Figsref}[1]{\ifnum\refstyle=1{Figs.~\ref{#1}}\else{figures~\ref{#1}}\fi}
\newcommand{\Appref}[1]{\ifnum\refstyle=1{App.~\ref{#1}}\else{appendix~\ref{#1}}\fi}
\newcommand{\Appsref}[1]{\ifnum\refstyle=1{Apps.~\ref{#1}}\else{appendices~\ref{#1}}\fi}
\newcommand{\Chref}[1]{\ifnum\refstyle=1{Ch.~\ref{#1}}\else{chapter~\ref{#1}}\fi}
\newcommand{\Chsref}[1]{\ifnum\refstyle=1{Chs.~\ref{#1}}\else{chapters~\ref{#1}}\fi}
\newcommand{\Secref}[1]{\ifnum\refstyle=1{Sec.~\ref{#1}}\else{section~\ref{#1}}\fi}
\newcommand{\Secsref}[1]{\ifnum\refstyle=1{Secs.~\ref{#1}}\else{sections~\ref{#1}}\fi}
\newcommand{\Refref}[1]{\ifnum\refstyle=1{Ref.~\cite{#1}}\else{reference~\cite{#1}}\fi}
\newcommand{\Refsref}[1]{\ifnum\refstyle=1{Refs.~\cite{#1}}\else{references~\cite{#1}}\fi}

\newcommand{\draftmode}{1}    
\newcommand{\notetoself}[1]{\ifnum \draftmode=1 {\color[rgb]{0,0,0.8} [#1]} \fi}  
\newcommand{\cuttext}[1]{\ifnum \draftmode=1 {\color[rgb]{0,0.5,0} [#1]} \fi}  
\newcommand{\warntext}[1]{{\ifnum \draftmode=1 \color[rgb]{0.9,0.6,0} \else  \color{black} \fi} #1}  
\usepackage{phonetic}
\pdfoutput=1

\renewcommand{\draftmode}{1} 
\newcommand{\Nor}{\ensuremath{\mathcal{N}}} 
\newcommand{\Nk}{\ensuremath{N}} 
\newcommand{\Eth}{\ensuremath{\mathcal{D}}}
\newcommand{\mdm}{\ensuremath{m_\mathrm{DM}}}  

\newcommand{\dm}{DM} 

\begin{document}


%
\title{Detecting Classically Undetectable Particles through Quantum Decoherence}
\date{\today}
\author{C.~Jess~Riedel}\
\affiliation{IBM Watson Research Center, Yorktown Heights, NY, USA}


\begin{abstract}
Some hypothetical particles are considered essentially undetectable because they are far too light and slow-moving to transfer appreciable energy or momentum to the normal matter that composes a detector.  I propose instead directly detecting such feeble particles, like sub-MeV dark matter or even gravitons, through their uniquely distinguishable decoherent effects on quantum devices like matter interferometers. More generally, decoherence can reveal phenomena that have arbitrarily little classical influence on normal matter, giving new motivation for the pursuit of macroscopic superpositions.
\end{abstract}

\maketitle

\begin{figure} [b!]
    \centering 
  \newcommand{\pbwidthfactor}{0.95}
    \includegraphics[width=\pbwidthfactor\columnwidth]{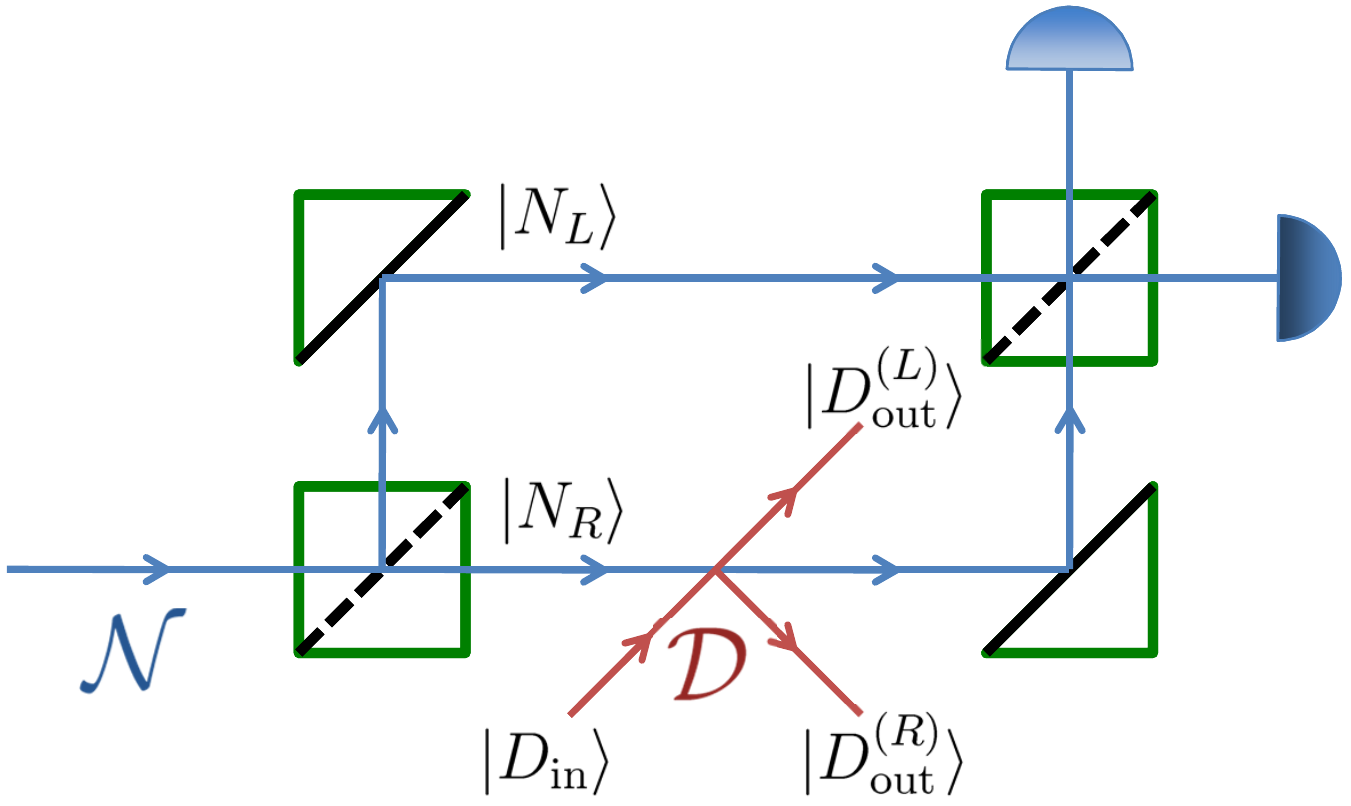}
  \caption{\textbf{Decoherence detection with a Mach-Zehnder interferometer.} System $\Nor$ is placed in a coherent superposition of spatially displaced wavepackets $\ket{\Nk_{L}}$ and $\ket{\Nk_{R}}$ that each travel a separate path and then are recombined.  In the absence of system $\Eth$, the interferometer is tuned so that $\Nor$ will be detected at the bright port with near unit probability, and at the dim port with near vanishing probability.  However, if system $\Eth$ scatters off $\Nor$, these two paths can decohere and $\Nor$ will be detected at the dim port 50\% of the time.}
  \label{mz_diagram}
\end{figure}


One limit of experimental physics is that it does not allow us to rule out the existence of new particles, forces, dimensions, or universes from which we are causally disconnected.  Even if some hypothetical new phenomenon has non-zero interactions with the well-known electrons and nucleons at our disposal, the coupling may always be so small that the influence on our equipment is negligible.  Experimenters are only able to rule out coupling strengths above some cutoff, and at some level of statistical significance.  There are many proposed and underway experimental searches for hypothetical new particles and forces (with varying degrees of theoretical and observational motivation) that, if they exist, interact only weakly with normal matter.  
Most notable is probably dark matter, but others include supersymmetric particles, new neutrino flavors, mirror matter, and fifth forces. 
In this letter, I propose searching for hypothetical new particles and forces by observing the quantum decoherence \cite{Zurek2003, SchlosshauerText} they cause rather than their classical influence. It turns out that this enables the observation of phenomena that are classically \emph{undetectable}.  I will highlight the search potential for two particles: sub-MeV dark matter, which will be visible to spaceborne matter interferometry experiments, and gravitons, which would be revealed by the coherent control of Planck-mass superpositions.

Consider a Mach-Zehnder atom interferometer which takes advantage of the de-Broglie-wave nature of matter, figure \ref{mz_diagram}.  A beam splitter prepares the center of mass of an atom $\Nor$ in an initial coherent superposition $\ket{\Nk_{L}} + \ket{\Nk_{R}}$ of two wavepackets, with one wavepacket taking the left path and one taking the right path.  After propagating over some distance for a time $T$, the packets are recombined and directed toward two sensors.  The splitters are aligned so that the sensors effectively measure $\Nor$ in the basis $\{\ket{\Nk_\pm} =  \ket{\Nk_{L}} \pm \ket{\Nk_{R}} \}$.  The outcome $\ket{\Nk_+}$ will be obtained with near unit probability.

Now we allow for the possibility of some hypothetical particle $\Eth$ that might be passing through the interferometer.  The state $\ket{D_\varnothing}$ represents the absence of $\Eth$ and we take the evolution to be trivial when it is not present:
\begin{align}
\Big[\ket{\Nk_{L}} + \ket{\Nk_{R}} \Big] \ket{D_\varnothing} \to \Big[\ket{\Nk_{L}} + \ket{\Nk_{R}} \Big] \ket{D_\varnothing} .
\end{align}
Measuring in the basis $\{\ket{\Nk_\pm} \}$ gives outcome $\ket{\Nk_+}$ with certainty, as expected.  But suppose the $\Eth$ particle approaches in state $\ket{D_\mathrm{in}}$ and decoheres the superposition by scattering off the atom,
\begin{align}
\Big[\ket{\Nk_{L}} + \ket{\Nk_{R}} \Big] \ket{D_\mathrm{in}} \to \ket{\Nk_{L}}\ket{D^{(L)}_\mathrm{out}} + \ket{\Nk_{R}} \ket{D^{(R)}_\mathrm{out}} ,
\end{align}
into the conditional states $\ket{D^{(L)}_\mathrm{out}}$ and $\ket{D^{(R)}_\mathrm{out}}$ with $\braket{D^{(L)}_\mathrm{out}}{D^{(R)}_\mathrm{out}} \approx 0$, thereby recording which-path information.  If the atom $\Nor$ is heavy enough compared to the $\Eth$ particle, the wavepackets $\ket{\Nk_{L}}$ and $\ket{\Nk_{R}}$ of the atom are not significantly perturbed following the scattering event.  But a measurement in the basis $\{\ket{\Nk_\pm} \}$ now gives outcome $\ket{\Nk_{-}}$ half of the time.  When it does, this is direct evidence for the existence of $\Eth$ \emph{even if it transfers negligible momentum to the atom}.

This basic idea is contained in the decoherence experiments of Hornberger et al.\ \cite{Hornberger2003b} and others, although it has never been suggested as a detection method.  In the Hornberger et al.\ experiment, coherent spatial superpositions of $\mathrm{C}_{70}$ fullerenes were demonstrated by passing them through several gratings and recording the interference pattern.  (This is a multi-slit near-field experiment rather than a Mach-Zehnder interferometer.)   The interference region was filled with a gas of molecules much smaller than the fullerenes, so that collisions recorded which-path information in the gas but only mildly deflected the fullerenes. The pressure was adjustable so that the presence of the gas---and moreover its density---could be inferred from the suppression of the interference fringes.  However, in that experiment the deflection of the fullerenes was not negligible; the increasing pressure exponentially suppressed the count rate in addition to the fringe visibility, so decoherence would have no advantage as a method of detection.


I will now start calling the $\Eth$ particle ``dark matter'' (\dm{}), although most of the ideas will apply more generally.  A single \dm{} scattering event may or may not fully decohere the atom $\Nor$, depending on whether the \dm{} de Broglie wavelength is small enough to resolve the separation of the wavepackets and yield orthogonal conditional out states of the \dm{} \cite{Joos1985}.  To calculate the pre-measurement state of the atom for the general case, we must sum the effects of the entire \dm{} flux the atom experiences as it passes through the interferometer.  The state of the atom  after a time $T$ is
\begin{align}
\rho_\Nor = \frac{1}{2}\begin{pmatrix} 1 & \gamma \\ \gamma^* & 1 \end{pmatrix}
\end{align}
in the $\{\ket{\Nk_L},  \ket{\Nk_R} \}$ basis.  Here, $\gamma = \exp [ - \int_0^T \!\! \dd t \,  F(\vec{\Delta x})]$ is the decoherence factor, $\vec{\Delta x}$ is the spatial displacement of the wavepackets, and $F(\vec{\Delta x})$ is a complex frequency calculable from the \dm{} flux and the nature of its scattering interaction \cite{riedel2013submittedPRD}. This is a case of the well-studied phenomena of collisional decoherence \cite{Joos1985, Gallis1990, Diosi1995, Hornberger2003a, Hornberger2006, Adler2006,SchlosshauerText}. 

The condition for effective decoherence is that $\abs{\gamma} \ll 1$, that is $\mathrm{Re} \, F(\vec{\Delta x}) \gtrsim 1/T$.  This diagonalizes the density matrix and drives the probabilities for activating the sensors at either arm of the interferometer both to 1/2.  But since collisions with \dm{} will be rare, there appears to be little chance of the atom being decohered by \dm{} during its short trip through the interferometer.

One way to increase the likelihood is to simply lengthen the arms of the interferometer or slow down the atom; the expected number of scattering events should be linear in exposure time $T$ for an uncorrelated flux of \dm{}.
But more powerfully, one can superpose clusters of many atoms.  That is, build a \emph{matter} interferometer with targets $\Nor$ that are as large as possible.  Each nucleon composing $\Nor$ can contribute an independent decoherence factor, effectively multiplying the 	decoherence rate $F_\mathrm{R} \equiv \mathrm{Re} \,F(\vec{\Delta x})$ by the total number of atoms.  

To achieve interference of large objects with ever smaller de Broglie wavelengths, modern time-domain interferometers can require a time interval proportional to the size of the object superposed \cite{Nimmrichter2011a, Haslinger2013}. (This is a testament to the difficulty of superposing large objects, but it also means that investing in larger masses yields big dividends.) In the case of \dm{} scattering from nucleons through the oft-studied spin-independent channel, there is additionally an enhancement due to \emph{coherent elastic scattering} that is also proportional to the superposed object size for sufficiently low \dm{} masses \cite{riedel2013submittedPRD}.  With these effects taken together,  the \dm{} sensitivity can scale like the \emph{cube} of the superposed object's quoted mass.   Happily, recent progress in matter interferometry has been stunning, with clear fringe patterns produced when interfering molecules composed of up to 430 atoms and in excess of 6,000 amu \cite{Gerlich2011}.  Future prospects are even stronger \cite{Nimmrichter2011a, Romero-Isart2011, asenbaum2013cavity}, and these have great potential for discovery. Techniques already being deployed \cite{Haslinger2013} are expected to achieve superpositions exceeding $10^6$ amu \cite{Nimmrichter2011a}.

Of course, there are many possible sources of decoherence; anomalous decoherence hardly implies the existence of new particles.  Still, note that the inverse statement \emph{is} true: the observation of interference effects (which establishes the existence of the superposition, and hence implies that all relevant sources of decoherence have been eliminated) implies that \dm{} has \emph{not} scattered (which sets robust upper bounds on the cross section $\sigma$).  Furthermore, if anomalous decoherence is observed, one can gather strong evidence that it is due to \dm{} by observing the functional dependence of the interference fringe visibility on experimental parameters.  For instance, the spatial extent $\Delta x$ of the superposition can be adjusted by separating the arms of an interferometer, while the exposure time $T$ can be changed by varying the speed of the target or lengthening the arms.  Depending on the design, varying the isotopic composition might allow one to adjust the \dm{} cross-section of the nuclei without affecting conventional (extra-nuclear) sources of decoherence. 

But the most striking evidence will come from modulating the incoming \dm{} flux itself \cite{riedel2013submittedPRD}. The \dm{} may be directly shielded from reaching the detector using normal materials, such as lead or concrete, for all of the parameter space considered below in figure \ref{fig:sigma}.  If interference fringe visibility is enhanced when adding shielding, and suppressed when removing it, this is direct evidence about the source of the decoherence. In more powerful experiments sensitive to lower cross sections, for which shielding is not feasible, natural variations due to the Earth's movement around the sun can instead be exploited.  Finally, the anisotropy of the \dm{} flux means that rotating the interferometer (i.e.\ the separation $\vec{\Delta x}$) changes the decoherence rate by a factor of order unity \cite{riedel2013submittedPRD}.  This naturally make interferometers \emph{directional} \dm{} detectors, which are known to be highly desirable \cite{Ahlenetal2010}, in part because they can give unmistakable evidence that the decoherence has galactic origins.

\begin{figure} [tb!]
  \centering 
\newcommand{\weirdfactor}{0.85}
\includegraphics[height=\weirdfactor\columnwidth]{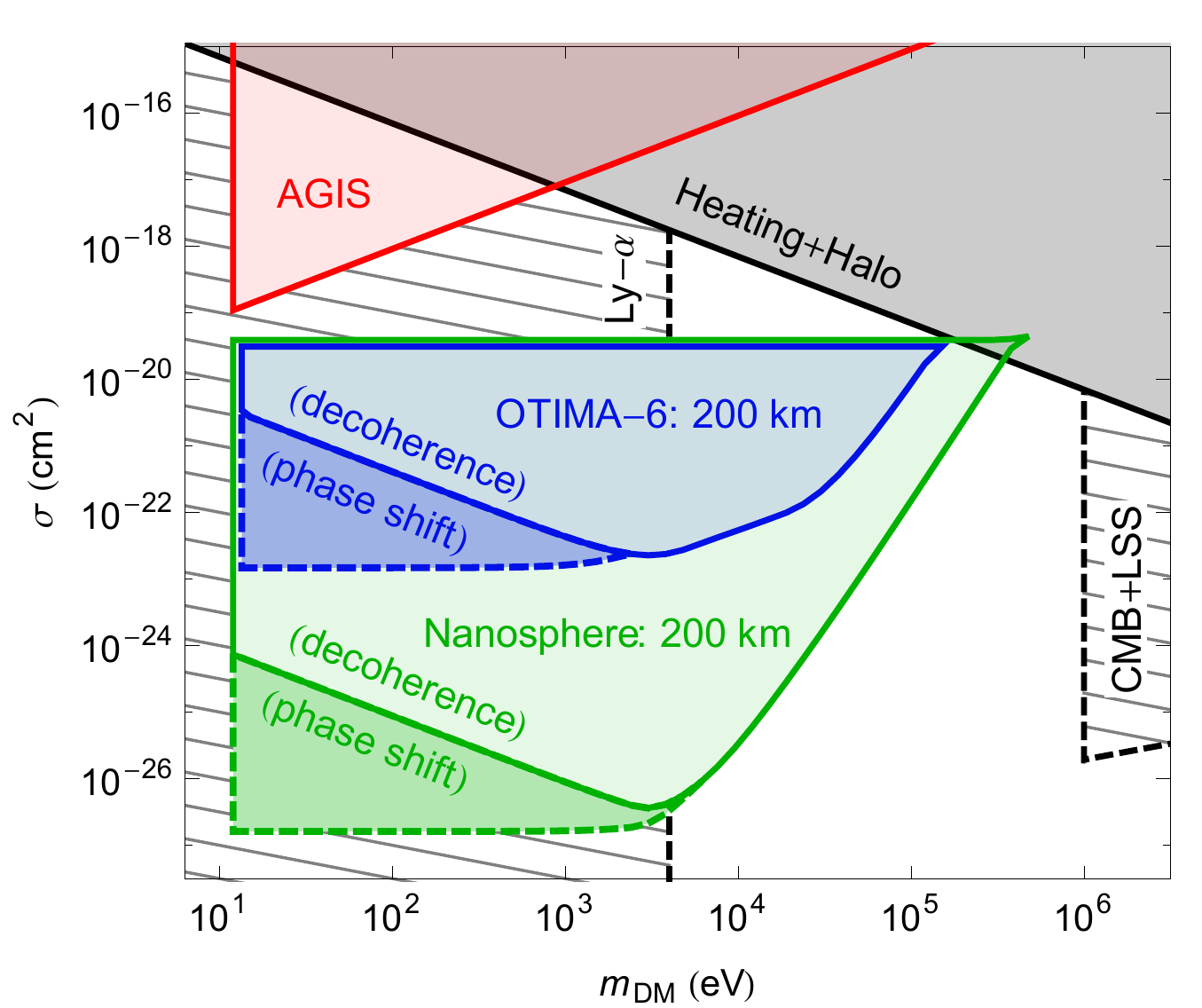}
\caption{\textbf{The sensitivity of some proposed superposition experiments to the spin-independent elastic scattering cross-section of dark matter with nucleons, compared with existing constraints.} The traditional distribution of \dm{} mass and velocity in the Earth's region of the Milky Way is assumed \cite{Catena2010, Lewin1996}.  The dark gray shaded region is already robustly excluded by heating and halo stability arguments in the Milky Way (``Heating+Halo'' \cite{Chivukula1990}).  Scenarios in which \dm{} is produced thermally in the early universe are incompatible with the hatched regions due to observations of the cosmic microwave background with large scale structure data (``CMB+LSS'' \cite{Chen2002}) and the Lyman-$\alpha$ forest (``Ly-$\alpha$'' \cite{Viel2008}).   Solid colored lines bound regions where \dm{} would cause decoherence in three proposed experiments: a satellite-based atom interferometer (``AGIS'' \cite{Dimopoulos2008}), 40 nm diameter optically-trapped silicon nanospheres (``Nanosphere'' \cite{Romero-Isart2011}), and the OTIMA interferometer with cluster of gold of mass $10^6$ amu (``OTIMA-$6$'' \cite{Nimmrichter2011a}). The border is defined by an $e$-fold suppression of the interference fringes:  $\abs{\gamma} = 1/e$. A successful AGIS satellite would set new exclusion limits on \dm{} where its sensitivity dips below the halo heating/stability bound for $\mdm \lesssim 1$ keV.  On the other hand, the OTIMA and nanosphere experiments would be shielded from \dm{} by the atmosphere if operated at sea level, so exclusion regions illustrate the sensitivity at an altitude of 200 km (reachable by sounding rocket or satellite).  The darker regions bordered by colored dashed lines indicates where the coherent phase shift could be observed without being overwhelmed by decoherence.  The separation vector $\vec{\Delta x}$ is taken to point into the \dm{} ``wind''. See Ref. \cite{riedel2013submittedPRD} for details.}
  \label{fig:sigma}
\end{figure} 


We expect our device to be useful when the momentum transfer during nuclear collisions is negligible, so I concentrate on the case where \dm{} is a particle with mass much smaller than the nucleon mass, say $\mdm \lesssim 1$ MeV$/c^2$.  (Indeed, direct-detection experiments searching for WIMPs are blind to such low masses.)  This means the scattering from nuclei should be effectively elastic because the \dm{} is far too feeble to excite internal nuclear states.  Additionally, the s-wave component of the partial-wave expansion is expected to dominate because of the very long de Broglie wavelength of sub-GeV \dm{} \cite{SquiresText}.

To demonstrate the potential in the future of detecting \dm{}-nucleon scattering through decoherence, I consider three experimental proposals currently being pursued: an optical time-domain ionizing matter-wave (OTIMA) interferometer that will interfere clusters of atoms larger than $10^6$ amu \cite{Nimmrichter2011a, Haslinger2013}, an optically-trapped 40 nm silicon nanosphere with mass $\sim \! \! 10^8$ amu \cite{Romero-Isart2011}, and the satellite-based Atomic Gravitational wave Interferometric Sensor (AGIS) \cite{Dimopoulos2008} using rubidium-87 atoms. Their sensitivity to \dm{} is depicted in figure \ref{fig:sigma}.  These experiments have been proposed for reasons that have nothing to do with discovering new particles; significant improvements in sensitivity are likely for devices designed with \dm{} in mind \cite{ArndtPC}.

Direct-detection experiments on Earth will only be sensitive to \dm{} if the scattering cross-section with nucleons is sufficiently low for \dm{} to pass through the atmosphere and reach the experiment.  This upper bounds the spin-independent cross-section visible to experiments on the Earth's surface at about $10^{-28.5} \mathrm{cm}^2$ (which the two terrestrial proposals cannot reach).  This can be circumvented by placing the experiment on a high-altitude balloon ($\sim \!\! 30$ km altitude; $\sim \!\! 10^{-26.5} \mathrm{cm}^2$), a sounding rocket ($\sim \!\! 200$ km altitude; $\sim\!\! 10^{-20.5} \mathrm{cm}^2$), or a satellite.  Note that the required weight to shield \dm{} in the range tested by balloon-, rocket-, and space-borne experiments is manageable \cite{riedel2013submittedPRD}. And in fact, orbital platforms for large quantum superposition are already under investigation because they offer several advantages, both for optical traps \cite{Kaltenbaek2012} and interferometers \cite{amelino-cameliaetal2008gauge, jentsch2004hyper, ertmeretal2009matter, arndt2013free-falling, Nimmrichter2011a}. The weightless environment features unlimited free-fall times and isolation from seismic vibration, increasing sensitivity by multiple orders of magnitude and allowing quantum superpositions not feasible on Earth \cite{Kaltenbaek2012, sorrentinoetal2010compact, amelino-cameliaetal2008gauge, jentsch2004hyper, ertmeretal2009matter}. 

Now let us briefly turn to gravitons, which are argued to be undetectable by any feasible classical measurement \cite{Rothman2006, Boughn2006, DysonReview}.  In particular, a Jupiter-sized detector (!) in close orbit around a neutron star is expected to absorb no more than one graviton every decade, and even then is unlikely to be able to distinguish it from background events in any imaginable manner \cite{Rothman2006}.  But a detector of decoherence may be able to reveal gravitons---although the technology for such a device still lies far in the future.

\begin{figure} [bt!]
    \centering 
  \newcommand{\pbwidthfactor}{0.95} \includegraphics[width=\pbwidthfactor\columnwidth]{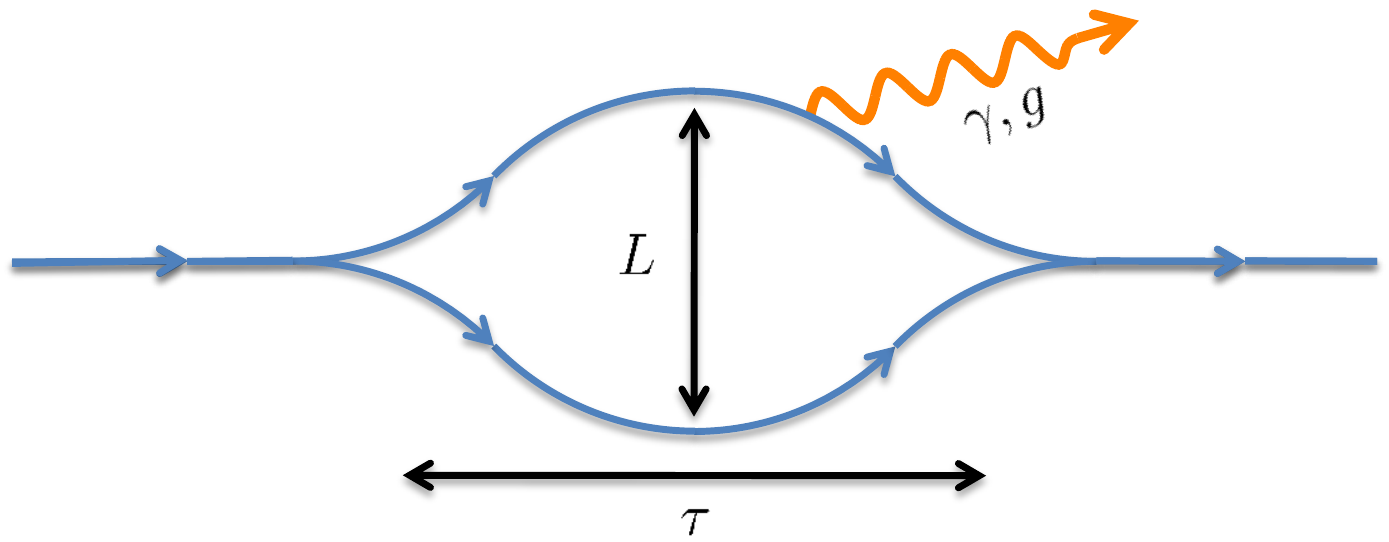}
  \caption{Decoherence by electromagnetic or gravitational bremsstrahlung. A charged object is brought into a coherent superposition of spatial extent $L$ then recombined after a time $\tau$.  If the two paths include sufficient relative acceleration for the given charge, they decohere through the emission of bremsstrahlung, which records with-path information.  The elementary charges or masses act together coherently to generate the radiation, which need not be absorbed for its existence to be detected.}
  \label{wiggler_diagram}
\end{figure}

Consider the toy matter interferometer in figure \ref{wiggler_diagram} in which a clump of mass $m$ is superposed over a distance $L$.  Even if there is no decoherence of the two paths from external environments, there can be substantial \emph{intrinsic} decoherence due to emitted radiation \cite{Breuer2001, BreuerBrem, Durr2000}.  When the clump is thermal, there are two types of sources: blackbody radiation and bremsstrahlung. 

Decoherence from blackbody radiation can be avoided by cooling the body to a temperature with characteristic wavelength much longer than $L$, and electromagnetic bremsstrahlung can be avoided by ensuring that there is no net electromagnetic charge.  But the gravitational charge (i.e.\ mass) is always positive, so gravitational bremsstrahlung is irreducible for paths with a given acceleration.  Gravity has no dipole radiation, unlike electromagnetism, so the primary contributor to bremsstrahlung is quadrupole radiation \cite{Gould1985, Weinberg1965}. The decoherence factor goes like $\gamma \sim \exp(- \alpha_\mathrm{G} \beta^4)$ where $\alpha_\mathrm{G} = G m^2/\hbar c$ is the effective gravitational coupling constant. (See the Supplementary methods.)  The Planck mass $m_\mathrm{P} \approx 21 \mu \mathrm{g} \approx 1.3 \times 10^{19}$ amu is precisely the mass scale at which $\alpha_\mathrm{G}$ reaches unity.  Thus, the coherent manipulation of Planck-mass superpositions at relativistic speeds will enable the detection of gravitons through decoherence. 

As mentioned, current matter interferometry technique are expected to achieve superpositions of masses exceeding $10^6$ amu \cite{Nimmrichter2011a}.  Spaceborne experiments should push this by multiple orders of magnitude using the same fundamental techniques \cite{Nimmrichter2011a, ArndtPC}.  Superpositions of lead spheres ($\sim 10^{14}$ amu) \cite{Romero-Isart2012} and of oscillating mirrors ($\sim 10^{16}$ amu) \cite{Marshall2003, Vitali2007} are being pursued, although the spatial extents of such superpositions are too small to decohere through bremsstrahlung.  Given this, the coherent manipulation of Planck-mass objects ($\sim 10^{19}$ amu)---though massively difficult---is not inconceivable.  Arguably, it is much more feasible than constructing a detector of Jovian proportions.

Of course, such a scheme for detecting the presence of gravitons assumes that all other sources of decoherence can be suppressed.  There are at least two irreducible backgrounds that, if they hinder the observation of gravitons, would themselves be exciting new physics: collisional decoherence from relic neutrinos, and bremsstrahlung from fifth forces stronger than gravity.  The investigation of these and other speculative possibilities is deferred to future work.


Beyond dark matter and gravitons, one can reinterpret many experiments (not just interferometers) that establish certain quantum states as direct evidence against hypothetical weak phenomena that, if existent, would decohere those states.    In principle, any superposition of matter states well separated in phase space is sensitive to collisional decoherence---and larger objects especially so.
Superposed mechanical oscillators \cite{OConnell2010, Romero-Isart2012, Pepper2012}, Bose-Einstein condensate interferometers \cite{Schumm2005}, and superconducting qubits \cite{clarke2008superconducting} have non-overlapping properties that may make them useful as detectors of decoherence caused by \dm{} or other new particles.  The toy Mach-Zehnder interferometer illustrates that the classical effects of such phenomena (e.g.\ momentum transfer) can be arbitrarily small while still causing very noticeable decoherence.

The essential difficulty in creating macroscopic superpositions is that the interaction of a single constituent particle is enough to decohere an arbitrarily large system, but this extreme sensitivity also gives them their detecting power.  Insofar as stability in the presence of decoherence defines the classicality of a quantum state \cite{Zurek1981, Zurek2003}, the best constraints on hypothetical weak phenomena will come from the most non-classical states. This gives new independent motivation for their experimental pursuit.

\subsection{Acknowledgements}

I thank Markus Arndt, Asimina Arvanitaki, Dirk Bouwmeester, Xiaoyong Chu, Savas Dimopoulos, Rouven Essig, Alexander Friedland, Jay Gambetta, Michael Graesser, Christian Hagmann, Steen Hannestad, Andrew Hime, Lorenzo Maccone, Gregory Mack, Jeremy Mardon, Benjamin Monreal, Kevin Moore, Harry Nelson, Bryon Neufeld, Shmuel Nussinov, Keith Rielage, Daniel Sank, Robert Scherrer, Alexia Schulz, Mark Srednicki, Paul Steinhardt, Lev Vaidman, Neal Weiner, and Haibo Yu for useful discussion.  I am especially grateful to Godfrey Miller for criticism, and to Charlie Bennett, Jim Hartle, and Wojciech Zurek for making this work possible. This research was partially supported by the U.S. Department of Energy through the LANL/LDRD program, and by the John Templeton Foundation through grant number 21484.

\bibliographystyle{apsrev4-1}
\bibliography{riedelbibetal}

\begin{thebibliography}{46}%
\makeatletter
\providecommand \@ifxundefined [1]{%
 \@ifx{#1\undefined}
}%
\providecommand \@ifnum [1]{%
 \ifnum #1\expandafter \@firstoftwo
 \else \expandafter \@secondoftwo
 \fi
}%
\providecommand \@ifx [1]{%
 \ifx #1\expandafter \@firstoftwo
 \else \expandafter \@secondoftwo
 \fi
}%
\providecommand \natexlab [1]{#1}%
\providecommand \enquote  [1]{``#1''}%
\providecommand \bibnamefont  [1]{#1}%
\providecommand \bibfnamefont [1]{#1}%
\providecommand \citenamefont [1]{#1}%
\providecommand \href@noop [0]{\@secondoftwo}%
\providecommand \href [0]{\begingroup \@sanitize@url \@href}%
\providecommand \@href[1]{\@@startlink{#1}\@@href}%
\providecommand \@@href[1]{\endgroup#1\@@endlink}%
\providecommand \@sanitize@url [0]{\catcode `\\12\catcode `\$12\catcode
  `\&12\catcode `\#12\catcode `\^12\catcode `\_12\catcode `\%12\relax}%
\providecommand \@@startlink[1]{}%
\providecommand \@@endlink[0]{}%
\providecommand \url  [0]{\begingroup\@sanitize@url \@url }%
\providecommand \@url [1]{\endgroup\@href {#1}{\urlprefix }}%
\providecommand \urlprefix  [0]{URL }%
\providecommand \Eprint [0]{\href }%
\providecommand \doibase [0]{http://dx.doi.org/}%
\providecommand \selectlanguage [0]{\@gobble}%
\providecommand \bibinfo  [0]{\@secondoftwo}%
\providecommand \bibfield  [0]{\@secondoftwo}%
\providecommand \translation [1]{[#1]}%
\providecommand \BibitemOpen [0]{}%
\providecommand \bibitemStop [0]{}%
\providecommand \bibitemNoStop [0]{.\EOS\space}%
\providecommand \EOS [0]{\spacefactor3000\relax}%
\providecommand \BibitemShut  [1]{\csname bibitem#1\endcsname}%
\let\auto@bib@innerbib\@empty
\bibitem [{\citenamefont {Zurek}(2003)}]{Zurek2003}%
  \BibitemOpen
  \bibfield  {author} {\bibinfo {author} {\bibfnamefont {W.~H.}\ \bibnamefont
  {Zurek}},\ }\href {\doibase 10.1103/RevModPhys.75.715} {\bibfield  {journal}
  {\bibinfo  {journal} {Rev. Mod. Phys.}\ }\textbf {\bibinfo {volume} {75}},\
  \bibinfo {pages} {715} (\bibinfo {year} {2003})}\BibitemShut {NoStop}%
\bibitem [{\citenamefont {Schlosshauer}(2008)}]{SchlosshauerText}%
  \BibitemOpen
  \bibfield  {author} {\bibinfo {author} {\bibfnamefont {M.}~\bibnamefont
  {Schlosshauer}},\ }\href@noop {} {\emph {\bibinfo {title} {Decoherence and
  the Quantum-to-Classical Transition}}}\ (\bibinfo  {publisher}
  {Springer-Verlag},\ \bibinfo {address} {Berlin},\ \bibinfo {year}
  {2008})\BibitemShut {NoStop}%
\bibitem [{\citenamefont {Hornberger}\ \emph {et~al.}(2003)\citenamefont
  {Hornberger}, \citenamefont {Uttenthaler}, \citenamefont {Brezger},
  \citenamefont {Hackerm\"uller}, \citenamefont {Arndt},\ and\ \citenamefont
  {Zeilinger}}]{Hornberger2003b}%
  \BibitemOpen
  \bibfield  {author} {\bibinfo {author} {\bibfnamefont {K.}~\bibnamefont
  {Hornberger}}, \bibinfo {author} {\bibfnamefont {S.}~\bibnamefont
  {Uttenthaler}}, \bibinfo {author} {\bibfnamefont {B.}~\bibnamefont
  {Brezger}}, \bibinfo {author} {\bibfnamefont {L.}~\bibnamefont
  {Hackerm\"uller}}, \bibinfo {author} {\bibfnamefont {M.}~\bibnamefont
  {Arndt}}, \ and\ \bibinfo {author} {\bibfnamefont {A.}~\bibnamefont
  {Zeilinger}},\ }\href {\doibase 10.1103/PhysRevLett.90.160401} {\bibfield
  {journal} {\bibinfo  {journal} {Phys. Rev. Lett.}\ }\textbf {\bibinfo
  {volume} {90}},\ \bibinfo {pages} {160401} (\bibinfo {year}
  {2003})}\BibitemShut {NoStop}%
\bibitem [{\citenamefont {Joos}\ and\ \citenamefont {Zeh}(1985)}]{Joos1985}%
  \BibitemOpen
  \bibfield  {author} {\bibinfo {author} {\bibfnamefont {E.}~\bibnamefont
  {Joos}}\ and\ \bibinfo {author} {\bibfnamefont {H.~D.}\ \bibnamefont {Zeh}},\
  }\href {\doibase 10.1007/BF01725541} {\bibfield  {journal} {\bibinfo
  {journal} {Zeitschrift f\"{u}r Physik B Condensed Matter}\ }\textbf {\bibinfo
  {volume} {59}},\ \bibinfo {pages} {223} (\bibinfo {year} {1985})}\BibitemShut
  {NoStop}%
\bibitem [{\citenamefont {Riedel}(2013)}]{riedel2013submittedPRD}%
  \BibitemOpen
  \bibfield  {author} {\bibinfo {author} {\bibfnamefont {C.~J.}\ \bibnamefont
  {Riedel}},\ }\href@noop {} {\enquote {\bibinfo {title} {Direct detection of
  classically undetectable dark matter through quantum decoherence},}\ }
  (\bibinfo {year} {2013}),\ \Eprint {http://arxiv.org/abs/Concurrent
  submission to \emph{Physical Review D}} {Concurrent submission to
  \emph{Physical Review D}} \BibitemShut {NoStop}%
\bibitem [{\citenamefont {Gallis}\ and\ \citenamefont
  {Fleming}(1990)}]{Gallis1990}%
  \BibitemOpen
  \bibfield  {author} {\bibinfo {author} {\bibfnamefont {M.~R.}\ \bibnamefont
  {Gallis}}\ and\ \bibinfo {author} {\bibfnamefont {G.~N.}\ \bibnamefont
  {Fleming}},\ }\href {\doibase 10.1103/PhysRevA.42.38} {\bibfield  {journal}
  {\bibinfo  {journal} {Phys. Rev. A}\ }\textbf {\bibinfo {volume} {42}},\
  \bibinfo {pages} {38} (\bibinfo {year} {1990})}\BibitemShut {NoStop}%
\bibitem [{\citenamefont {Diosi}(1995)}]{Diosi1995}%
  \BibitemOpen
  \bibfield  {author} {\bibinfo {author} {\bibfnamefont {L.}~\bibnamefont
  {Diosi}},\ }\href {\doibase 10.1209/0295-5075/30/2/001} {\bibfield  {journal}
  {\bibinfo  {journal} {Europhys. Lett.}\ }\textbf {\bibinfo {volume} {30}},\
  \bibinfo {pages} {63} (\bibinfo {year} {1995})}\BibitemShut {NoStop}%
\bibitem [{\citenamefont {Hornberger}\ and\ \citenamefont
  {Sipe}(2003)}]{Hornberger2003a}%
  \BibitemOpen
  \bibfield  {author} {\bibinfo {author} {\bibfnamefont {K.}~\bibnamefont
  {Hornberger}}\ and\ \bibinfo {author} {\bibfnamefont {J.~E.}\ \bibnamefont
  {Sipe}},\ }\href {\doibase 10.1103/PhysRevA.68.012105} {\bibfield  {journal}
  {\bibinfo  {journal} {Phys. Rev. A}\ }\textbf {\bibinfo {volume} {68}},\
  \bibinfo {pages} {012105} (\bibinfo {year} {2003})}\BibitemShut {NoStop}%
\bibitem [{\citenamefont {Hornberger}(2006)}]{Hornberger2006}%
  \BibitemOpen
  \bibfield  {author} {\bibinfo {author} {\bibfnamefont {K.}~\bibnamefont
  {Hornberger}},\ }\href {\doibase 10.1103/PhysRevLett.97.060601} {\bibfield
  {journal} {\bibinfo  {journal} {Phys. Rev. Lett.}\ }\textbf {\bibinfo
  {volume} {97}},\ \bibinfo {pages} {060601} (\bibinfo {year}
  {2006})}\BibitemShut {NoStop}%
\bibitem [{\citenamefont {Adler}(2006)}]{Adler2006}%
  \BibitemOpen
  \bibfield  {author} {\bibinfo {author} {\bibfnamefont {S.~L.}\ \bibnamefont
  {Adler}},\ }\href@noop {} {\bibfield  {journal} {\bibinfo  {journal} {Journal
  of Physics A: Mathematical and General}\ }\textbf {\bibinfo {volume} {39}},\
  \bibinfo {pages} {14067} (\bibinfo {year} {2006})}\BibitemShut {NoStop}%
\bibitem [{\citenamefont {Nimmrichter}\ \emph {et~al.}(2011)\citenamefont
  {Nimmrichter}, \citenamefont {Haslinger}, \citenamefont {Hornberger},\ and\
  \citenamefont {Arndt}}]{Nimmrichter2011a}%
  \BibitemOpen
  \bibfield  {author} {\bibinfo {author} {\bibfnamefont {S.}~\bibnamefont
  {Nimmrichter}}, \bibinfo {author} {\bibfnamefont {P.}~\bibnamefont
  {Haslinger}}, \bibinfo {author} {\bibfnamefont {K.}~\bibnamefont
  {Hornberger}}, \ and\ \bibinfo {author} {\bibfnamefont {M.}~\bibnamefont
  {Arndt}},\ }\href {http://stacks.iop.org/1367-2630/13/i=7/a=075002}
  {\bibfield  {journal} {\bibinfo  {journal} {New Journal of Physics}\ }\textbf
  {\bibinfo {volume} {13}},\ \bibinfo {pages} {075002} (\bibinfo {year}
  {2011})}\BibitemShut {NoStop}%
\bibitem [{\citenamefont {Haslinger}\ \emph {et~al.}(2013)\citenamefont
  {Haslinger}, \citenamefont {D{\"o}rre}, \citenamefont {Geyer}, \citenamefont
  {Rodewald}, \citenamefont {Nimmrichter},\ and\ \citenamefont
  {Arndt}}]{Haslinger2013}%
  \BibitemOpen
  \bibfield  {author} {\bibinfo {author} {\bibfnamefont {P.}~\bibnamefont
  {Haslinger}}, \bibinfo {author} {\bibfnamefont {N.}~\bibnamefont
  {D{\"o}rre}}, \bibinfo {author} {\bibfnamefont {P.}~\bibnamefont {Geyer}},
  \bibinfo {author} {\bibfnamefont {J.}~\bibnamefont {Rodewald}}, \bibinfo
  {author} {\bibfnamefont {S.}~\bibnamefont {Nimmrichter}}, \ and\ \bibinfo
  {author} {\bibfnamefont {M.}~\bibnamefont {Arndt}},\ }\href@noop {}
  {\bibfield  {journal} {\bibinfo  {journal} {Nature Physics}\ } (\bibinfo
  {year} {2013})}\BibitemShut {NoStop}%
\bibitem [{\citenamefont {Gerlich}\ \emph {et~al.}(2011)\citenamefont
  {Gerlich}, \citenamefont {Eibenberger}, \citenamefont {Tomandl},
  \citenamefont {Nimmrichter}, \citenamefont {Hornberger}, \citenamefont
  {Fagan}, \citenamefont {T{\"u}xen}, \citenamefont {Mayor},\ and\
  \citenamefont {Arndt}}]{Gerlich2011}%
  \BibitemOpen
  \bibfield  {author} {\bibinfo {author} {\bibfnamefont {S.}~\bibnamefont
  {Gerlich}}, \bibinfo {author} {\bibfnamefont {S.}~\bibnamefont
  {Eibenberger}}, \bibinfo {author} {\bibfnamefont {M.}~\bibnamefont
  {Tomandl}}, \bibinfo {author} {\bibfnamefont {S.}~\bibnamefont
  {Nimmrichter}}, \bibinfo {author} {\bibfnamefont {K.}~\bibnamefont
  {Hornberger}}, \bibinfo {author} {\bibfnamefont {P.~J.}\ \bibnamefont
  {Fagan}}, \bibinfo {author} {\bibfnamefont {J.}~\bibnamefont {T{\"u}xen}},
  \bibinfo {author} {\bibfnamefont {M.}~\bibnamefont {Mayor}}, \ and\ \bibinfo
  {author} {\bibfnamefont {M.}~\bibnamefont {Arndt}},\ }\href
  {http://dx.doi.org/10.1038/ncomms1263} {\bibfield  {journal} {\bibinfo
  {journal} {Nat Commun}\ }\textbf {\bibinfo {volume} {2}} (\bibinfo {year}
  {2011})}\BibitemShut {NoStop}%
\bibitem [{\citenamefont {Romero-Isart}\ \emph {et~al.}(2011)\citenamefont
  {Romero-Isart}, \citenamefont {Pflanzer}, \citenamefont {Blaser},
  \citenamefont {Kaltenbaek}, \citenamefont {Kiesel}, \citenamefont
  {Aspelmeyer},\ and\ \citenamefont {Cirac}}]{Romero-Isart2011}%
  \BibitemOpen
  \bibfield  {author} {\bibinfo {author} {\bibfnamefont {O.}~\bibnamefont
  {Romero-Isart}}, \bibinfo {author} {\bibfnamefont {A.~C.}\ \bibnamefont
  {Pflanzer}}, \bibinfo {author} {\bibfnamefont {F.}~\bibnamefont {Blaser}},
  \bibinfo {author} {\bibfnamefont {R.}~\bibnamefont {Kaltenbaek}}, \bibinfo
  {author} {\bibfnamefont {N.}~\bibnamefont {Kiesel}}, \bibinfo {author}
  {\bibfnamefont {M.}~\bibnamefont {Aspelmeyer}}, \ and\ \bibinfo {author}
  {\bibfnamefont {J.~I.}\ \bibnamefont {Cirac}},\ }\href {\doibase
  10.1103/PhysRevLett.107.020405} {\bibfield  {journal} {\bibinfo  {journal}
  {Phys. Rev. Lett.}\ }\textbf {\bibinfo {volume} {107}},\ \bibinfo {pages}
  {020405} (\bibinfo {year} {2011})}\BibitemShut {NoStop}%
\bibitem [{\citenamefont {Asenbaum}\ \emph {et~al.}(2013)\citenamefont
  {Asenbaum}, \citenamefont {Kuhn}, \citenamefont {Nimmrichter}, \citenamefont
  {Sezer},\ and\ \citenamefont {Arndt}}]{asenbaum2013cavity}%
  \BibitemOpen
  \bibfield  {author} {\bibinfo {author} {\bibfnamefont {P.}~\bibnamefont
  {Asenbaum}}, \bibinfo {author} {\bibfnamefont {S.}~\bibnamefont {Kuhn}},
  \bibinfo {author} {\bibfnamefont {S.}~\bibnamefont {Nimmrichter}}, \bibinfo
  {author} {\bibfnamefont {U.}~\bibnamefont {Sezer}}, \ and\ \bibinfo {author}
  {\bibfnamefont {M.}~\bibnamefont {Arndt}},\ }\href@noop {} {\enquote
  {\bibinfo {title} {Cavity cooling of free silicon nanoparticles in
  high-vacuum},}\ } (\bibinfo {year} {2013}),\ \Eprint
  {http://arxiv.org/abs/arXiv:1306.4617} {arXiv:1306.4617} \BibitemShut
  {NoStop}%
\bibitem [{\citenamefont {Ahlen}\ \emph {et~al.}(2010)\citenamefont {Ahlen}
  \emph {et~al.}}]{Ahlenetal2010}%
  \BibitemOpen
  \bibfield  {author} {\bibinfo {author} {\bibfnamefont {S.}~\bibnamefont
  {Ahlen}} \emph {et~al.},\ }\href@noop {} {\bibfield  {journal} {\bibinfo
  {journal} {International Journal of Modern Physics A}\ }\textbf {\bibinfo
  {volume} {25}},\ \bibinfo {pages} {1} (\bibinfo {year} {2010})}\BibitemShut
  {NoStop}%
\bibitem [{\citenamefont {Catena}\ and\ \citenamefont
  {Ullio}(2010)}]{Catena2010}%
  \BibitemOpen
  \bibfield  {author} {\bibinfo {author} {\bibfnamefont {R.}~\bibnamefont
  {Catena}}\ and\ \bibinfo {author} {\bibfnamefont {P.}~\bibnamefont {Ullio}},\
  }\href {http://stacks.iop.org/1475-7516/2010/i=08/a=004} {\bibfield
  {journal} {\bibinfo  {journal} {Journal of Cosmology and Astroparticle
  Physics}\ }\textbf {\bibinfo {volume} {2010}},\ \bibinfo {pages} {004}
  (\bibinfo {year} {2010})}\BibitemShut {NoStop}%
\bibitem [{\citenamefont {Lewin}\ and\ \citenamefont
  {Smith}(1996)}]{Lewin1996}%
  \BibitemOpen
  \bibfield  {author} {\bibinfo {author} {\bibfnamefont {J.}~\bibnamefont
  {Lewin}}\ and\ \bibinfo {author} {\bibfnamefont {P.}~\bibnamefont {Smith}},\
  }\href {\doibase 10.1016/S0927-6505(96)00047-3} {\bibfield  {journal}
  {\bibinfo  {journal} {Astroparticle Physics}\ }\textbf {\bibinfo {volume}
  {6}},\ \bibinfo {pages} {87 } (\bibinfo {year} {1996})}\BibitemShut {NoStop}%
\bibitem [{\citenamefont {Chivukula}\ \emph {et~al.}(1990)\citenamefont
  {Chivukula}, \citenamefont {Cohen}, \citenamefont {Dimopoulos},\ and\
  \citenamefont {Walker}}]{Chivukula1990}%
  \BibitemOpen
  \bibfield  {author} {\bibinfo {author} {\bibfnamefont {R.~S.}\ \bibnamefont
  {Chivukula}}, \bibinfo {author} {\bibfnamefont {A.~G.}\ \bibnamefont
  {Cohen}}, \bibinfo {author} {\bibfnamefont {S.}~\bibnamefont {Dimopoulos}}, \
  and\ \bibinfo {author} {\bibfnamefont {T.~P.}\ \bibnamefont {Walker}},\
  }\href {\doibase 10.1103/PhysRevLett.65.957} {\bibfield  {journal} {\bibinfo
  {journal} {Phys. Rev. Lett.}\ }\textbf {\bibinfo {volume} {65}},\ \bibinfo
  {pages} {957} (\bibinfo {year} {1990})}\BibitemShut {NoStop}%
\bibitem [{\citenamefont {Chen}\ \emph {et~al.}(2002)\citenamefont {Chen},
  \citenamefont {Hannestad},\ and\ \citenamefont {Scherrer}}]{Chen2002}%
  \BibitemOpen
  \bibfield  {author} {\bibinfo {author} {\bibfnamefont {X.}~\bibnamefont
  {Chen}}, \bibinfo {author} {\bibfnamefont {S.}~\bibnamefont {Hannestad}}, \
  and\ \bibinfo {author} {\bibfnamefont {R.~J.}\ \bibnamefont {Scherrer}},\
  }\href {\doibase 10.1103/PhysRevD.65.123515} {\bibfield  {journal} {\bibinfo
  {journal} {Phys. Rev. D}\ }\textbf {\bibinfo {volume} {65}},\ \bibinfo
  {pages} {123515} (\bibinfo {year} {2002})}\BibitemShut {NoStop}%
\bibitem [{\citenamefont {Viel}\ \emph {et~al.}(2008)\citenamefont {Viel},
  \citenamefont {Becker}, \citenamefont {Bolton}, \citenamefont {Haehnelt},
  \citenamefont {Rauch},\ and\ \citenamefont {Sargent}}]{Viel2008}%
  \BibitemOpen
  \bibfield  {author} {\bibinfo {author} {\bibfnamefont {M.}~\bibnamefont
  {Viel}}, \bibinfo {author} {\bibfnamefont {G.~D.}\ \bibnamefont {Becker}},
  \bibinfo {author} {\bibfnamefont {J.~S.}\ \bibnamefont {Bolton}}, \bibinfo
  {author} {\bibfnamefont {M.~G.}\ \bibnamefont {Haehnelt}}, \bibinfo {author}
  {\bibfnamefont {M.}~\bibnamefont {Rauch}}, \ and\ \bibinfo {author}
  {\bibfnamefont {W.~L.}\ \bibnamefont {Sargent}},\ }\href@noop {} {\bibfield
  {journal} {\bibinfo  {journal} {Physical Review Letters}\ }\textbf {\bibinfo
  {volume} {100}},\ \bibinfo {pages} {041304} (\bibinfo {year}
  {2008})}\BibitemShut {NoStop}%
\bibitem [{\citenamefont {Dimopoulos}\ \emph {et~al.}(2008)\citenamefont
  {Dimopoulos}, \citenamefont {Graham}, \citenamefont {Hogan}, \citenamefont
  {Kasevich},\ and\ \citenamefont {Rajendran}}]{Dimopoulos2008}%
  \BibitemOpen
  \bibfield  {author} {\bibinfo {author} {\bibfnamefont {S.}~\bibnamefont
  {Dimopoulos}}, \bibinfo {author} {\bibfnamefont {P.~W.}\ \bibnamefont
  {Graham}}, \bibinfo {author} {\bibfnamefont {J.~M.}\ \bibnamefont {Hogan}},
  \bibinfo {author} {\bibfnamefont {M.~A.}\ \bibnamefont {Kasevich}}, \ and\
  \bibinfo {author} {\bibfnamefont {S.}~\bibnamefont {Rajendran}},\ }\href
  {\doibase 10.1103/PhysRevD.78.122002} {\bibfield  {journal} {\bibinfo
  {journal} {Phys. Rev. D}\ }\textbf {\bibinfo {volume} {78}},\ \bibinfo
  {pages} {122002} (\bibinfo {year} {2008})}\BibitemShut {NoStop}%
\bibitem [{\citenamefont {Squires}(1978)}]{SquiresText}%
  \BibitemOpen
  \bibfield  {author} {\bibinfo {author} {\bibfnamefont {G.~L.}\ \bibnamefont
  {Squires}},\ }\href@noop {} {\emph {\bibinfo {title} {Introduction to the
  Theory of Thermal Neutron Scattering}}}\ (\bibinfo  {publisher} {Cambridge
  University Press},\ \bibinfo {address} {New York},\ \bibinfo {year}
  {1978})\BibitemShut {NoStop}%
\bibitem [{\citenamefont {Arndt}(2012)}]{ArndtPC}%
  \BibitemOpen
  \bibfield  {author} {\bibinfo {author} {\bibfnamefont {M.}~\bibnamefont
  {Arndt}},\ }\href@noop {} {} (\bibinfo {year} {2012}),\ \bibinfo {note}
  {private communication}\BibitemShut {NoStop}%
\bibitem [{\citenamefont {Kaltenbaek}\ \emph {et~al.}(2012)\citenamefont
  {Kaltenbaek}, \citenamefont {Hechenblaikner}, \citenamefont {Kiesel},
  \citenamefont {Romero-Isart}, \citenamefont {Schwab}, \citenamefont
  {Johann},\ and\ \citenamefont {Aspelmeyer}}]{Kaltenbaek2012}%
  \BibitemOpen
  \bibfield  {author} {\bibinfo {author} {\bibfnamefont {R.}~\bibnamefont
  {Kaltenbaek}}, \bibinfo {author} {\bibfnamefont {G.}~\bibnamefont
  {Hechenblaikner}}, \bibinfo {author} {\bibfnamefont {N.}~\bibnamefont
  {Kiesel}}, \bibinfo {author} {\bibfnamefont {O.}~\bibnamefont
  {Romero-Isart}}, \bibinfo {author} {\bibfnamefont {K.}~\bibnamefont
  {Schwab}}, \bibinfo {author} {\bibfnamefont {U.}~\bibnamefont {Johann}}, \
  and\ \bibinfo {author} {\bibfnamefont {M.}~\bibnamefont {Aspelmeyer}},\
  }\href@noop {} {\bibfield  {journal} {\bibinfo  {journal} {Experimental
  Astronomy}\ ,\ \bibinfo {pages} {1}} (\bibinfo {year} {2012})}\BibitemShut
  {NoStop}%
\bibitem [{\citenamefont {Amelino-Camelia}\ \emph {et~al.}(2008)\citenamefont
  {Amelino-Camelia} \emph {et~al.}}]{amelino-cameliaetal2008gauge}%
  \BibitemOpen
  \bibfield  {author} {\bibinfo {author} {\bibfnamefont {G.}~\bibnamefont
  {Amelino-Camelia}} \emph {et~al.},\ }\href {\doibase
  10.1007/s10686-008-9086-9} {\bibfield  {journal} {\bibinfo  {journal}
  {Experimental Astronomy}\ }\textbf {\bibinfo {volume} {23}},\ \bibinfo
  {pages} {549} (\bibinfo {year} {2008})}\BibitemShut {NoStop}%
\bibitem [{\citenamefont {Jentsch}\ \emph {et~al.}(2004)\citenamefont
  {Jentsch}, \citenamefont {Müller}, \citenamefont {Rasel},\ and\ \citenamefont
  {Ertmer}}]{jentsch2004hyper}%
  \BibitemOpen
  \bibfield  {author} {\bibinfo {author} {\bibfnamefont {C.}~\bibnamefont
  {Jentsch}}, \bibinfo {author} {\bibfnamefont {T.}~\bibnamefont {Müller}},
  \bibinfo {author} {\bibfnamefont {E.~M.}\ \bibnamefont {Rasel}}, \ and\
  \bibinfo {author} {\bibfnamefont {W.}~\bibnamefont {Ertmer}},\ }\href
  {\doibase 10.1023/B:GERG.0000046179.26175.fa} {\bibfield  {journal} {\bibinfo
   {journal} {General Relativity and Gravitation}\ }\textbf {\bibinfo {volume}
  {36}},\ \bibinfo {pages} {2197} (\bibinfo {year} {2004})}\BibitemShut
  {NoStop}%
\bibitem [{\citenamefont {Ertmer}\ \emph {et~al.}(2009)\citenamefont {Ertmer}
  \emph {et~al.}}]{ertmeretal2009matter}%
  \BibitemOpen
  \bibfield  {author} {\bibinfo {author} {\bibfnamefont {W.}~\bibnamefont
  {Ertmer}} \emph {et~al.},\ }\href@noop {} {\bibfield  {journal} {\bibinfo
  {journal} {Experimental Astronomy}\ }\textbf {\bibinfo {volume} {23}},\
  \bibinfo {pages} {611} (\bibinfo {year} {2009})}\BibitemShut {NoStop}%
\bibitem [{\citenamefont {Arndt}(2013)}]{arndt2013free-falling}%
  \BibitemOpen
  \bibfield  {author} {\bibinfo {author} {\bibfnamefont {M.}~\bibnamefont
  {Arndt}},\ }\href {\doibase 10.1103/Physics.6.23} {\bibfield  {journal}
  {\bibinfo  {journal} {Physics}\ }\textbf {\bibinfo {volume} {6}},\ \bibinfo
  {pages} {23} (\bibinfo {year} {2013})}\BibitemShut {NoStop}%
\bibitem [{\citenamefont {Sorrentino}\ \emph {et~al.}(2010)\citenamefont
  {Sorrentino} \emph {et~al.}}]{sorrentinoetal2010compact}%
  \BibitemOpen
  \bibfield  {author} {\bibinfo {author} {\bibfnamefont {F.}~\bibnamefont
  {Sorrentino}} \emph {et~al.},\ }\href {\doibase 10.1007/s12217-010-9240-7}
  {\bibfield  {journal} {\bibinfo  {journal} {Microgravity Science and
  Technology}\ }\textbf {\bibinfo {volume} {22}},\ \bibinfo {pages} {551}
  (\bibinfo {year} {2010})}\BibitemShut {NoStop}%
\bibitem [{\citenamefont {Rothman}\ and\ \citenamefont
  {Boughn}(2006)}]{Rothman2006}%
  \BibitemOpen
  \bibfield  {author} {\bibinfo {author} {\bibfnamefont {T.}~\bibnamefont
  {Rothman}}\ and\ \bibinfo {author} {\bibfnamefont {S.}~\bibnamefont
  {Boughn}},\ }\href {http://dx.doi.org/10.1007/s10701-006-9081-9} {\bibfield
  {journal} {\bibinfo  {journal} {Foundations of Physics}\ }\textbf {\bibinfo
  {volume} {36}},\ \bibinfo {pages} {1801} (\bibinfo {year} {2006})},\ \bibinfo
  {note} {10.1007/s10701-006-9081-9}\BibitemShut {NoStop}%
\bibitem [{\citenamefont {Boughn}\ and\ \citenamefont
  {Rothman}(2006)}]{Boughn2006}%
  \BibitemOpen
  \bibfield  {author} {\bibinfo {author} {\bibfnamefont {S.}~\bibnamefont
  {Boughn}}\ and\ \bibinfo {author} {\bibfnamefont {T.}~\bibnamefont
  {Rothman}},\ }\href {http://stacks.iop.org/0264-9381/23/i=20/a=006}
  {\bibfield  {journal} {\bibinfo  {journal} {Classical and Quantum Gravity}\
  }\textbf {\bibinfo {volume} {23}},\ \bibinfo {pages} {5839} (\bibinfo {year}
  {2006})}\BibitemShut {NoStop}%
\bibitem [{\citenamefont {Dyson}(2004)}]{DysonReview}%
  \BibitemOpen
  \bibfield  {author} {\bibinfo {author} {\bibfnamefont {F.~J.}\ \bibnamefont
  {Dyson}},\ }in\ \href@noop {} {\emph {\bibinfo {booktitle} {NY Rev. Books 51
  (8)}}},\ \bibinfo {editor} {edited by\ \bibinfo {editor} {\bibfnamefont
  {M.}~\bibnamefont {Kafatos}}}\ (\bibinfo {year} {2004})\BibitemShut {NoStop}%
\bibitem [{\citenamefont {Breuer}\ and\ \citenamefont
  {Petruccione}(2001)}]{Breuer2001}%
  \BibitemOpen
  \bibfield  {author} {\bibinfo {author} {\bibfnamefont {H.-P.}\ \bibnamefont
  {Breuer}}\ and\ \bibinfo {author} {\bibfnamefont {F.}~\bibnamefont
  {Petruccione}},\ }\href {\doibase 10.1103/PhysRevA.63.032102} {\bibfield
  {journal} {\bibinfo  {journal} {Phys. Rev. A}\ }\textbf {\bibinfo {volume}
  {63}},\ \bibinfo {pages} {032102} (\bibinfo {year} {2001})}\BibitemShut
  {NoStop}%
\bibitem [{\citenamefont {Breuer}\ and\ \citenamefont
  {Petruccione}(2002)}]{BreuerBrem}%
  \BibitemOpen
  \bibfield  {author} {\bibinfo {author} {\bibfnamefont {H.-P.}\ \bibnamefont
  {Breuer}}\ and\ \bibinfo {author} {\bibfnamefont {F.}~\bibnamefont
  {Petruccione}},\ }\enquote {\bibinfo {title} {The theory of open quantum
  systems},}\ \ (\bibinfo  {publisher} {Oxford University Press},\ \bibinfo
  {address} {USA},\ \bibinfo {year} {2002})\ pp.\ \bibinfo {pages}
  {577--605}\BibitemShut {NoStop}%
\bibitem [{\citenamefont {D{\"u}rr}\ and\ \citenamefont
  {Spohn}(2000)}]{Durr2000}%
  \BibitemOpen
  \bibfield  {author} {\bibinfo {author} {\bibfnamefont {D.}~\bibnamefont
  {D{\"u}rr}}\ and\ \bibinfo {author} {\bibfnamefont {H.}~\bibnamefont
  {Spohn}},\ }in\ \href {http://dx.doi.org/10.1007/3-540-46657-6_6} {\emph
  {\bibinfo {booktitle} {Decoherence: Theoretical, Experimental, and Conceptual
  Problems}}},\ \bibinfo {series} {Lecture Notes in Physics}, Vol.\ \bibinfo
  {volume} {538},\ \bibinfo {editor} {edited by\ \bibinfo {editor}
  {\bibfnamefont {P.}~\bibnamefont {Blanchard}}, \bibinfo {editor}
  {\bibfnamefont {E.}~\bibnamefont {Joos}}, \bibinfo {editor} {\bibfnamefont
  {D.}~\bibnamefont {Giulini}}, \bibinfo {editor} {\bibfnamefont
  {C.}~\bibnamefont {Kiefer}}, \ and\ \bibinfo {editor} {\bibfnamefont {I.-O.}\
  \bibnamefont {Stamatescu}}}\ (\bibinfo  {publisher} {Springer Berlin /
  Heidelberg},\ \bibinfo {year} {2000})\ pp.\ \bibinfo {pages}
  {77--86}\BibitemShut {NoStop}%
\bibitem [{\citenamefont {Gould}(1985)}]{Gould1985}%
  \BibitemOpen
  \bibfield  {author} {\bibinfo {author} {\bibfnamefont {R.~J.}\ \bibnamefont
  {Gould}},\ }\href@noop {} {\bibfield  {journal} {\bibinfo  {journal} {The
  Astrophysical Journal}\ }\textbf {\bibinfo {volume} {288}},\ \bibinfo {pages}
  {789} (\bibinfo {year} {1985})}\BibitemShut {NoStop}%
\bibitem [{\citenamefont {Weinberg}(1965)}]{Weinberg1965}%
  \BibitemOpen
  \bibfield  {author} {\bibinfo {author} {\bibfnamefont {S.}~\bibnamefont
  {Weinberg}},\ }\href {\doibase 10.1103/PhysRev.140.B516} {\bibfield
  {journal} {\bibinfo  {journal} {Phys. Rev.}\ }\textbf {\bibinfo {volume}
  {140}},\ \bibinfo {pages} {B516} (\bibinfo {year} {1965})}\BibitemShut
  {NoStop}%
\bibitem [{\citenamefont {Romero-Isart}\ \emph {et~al.}(2012)\citenamefont
  {Romero-Isart}, \citenamefont {Clemente}, \citenamefont {Navau},
  \citenamefont {Sanchez},\ and\ \citenamefont {Cirac}}]{Romero-Isart2012}%
  \BibitemOpen
  \bibfield  {author} {\bibinfo {author} {\bibfnamefont {O.}~\bibnamefont
  {Romero-Isart}}, \bibinfo {author} {\bibfnamefont {L.}~\bibnamefont
  {Clemente}}, \bibinfo {author} {\bibfnamefont {C.}~\bibnamefont {Navau}},
  \bibinfo {author} {\bibfnamefont {A.}~\bibnamefont {Sanchez}}, \ and\
  \bibinfo {author} {\bibfnamefont {J.~I.}\ \bibnamefont {Cirac}},\ }\href
  {\doibase 10.1103/PhysRevLett.109.147205} {\bibfield  {journal} {\bibinfo
  {journal} {Phys. Rev. Lett.}\ }\textbf {\bibinfo {volume} {109}},\ \bibinfo
  {pages} {147205} (\bibinfo {year} {2012})}\BibitemShut {NoStop}%
\bibitem [{\citenamefont {Marshall}\ \emph {et~al.}(2003)\citenamefont
  {Marshall}, \citenamefont {Simon}, \citenamefont {Penrose},\ and\
  \citenamefont {Bouwmeester}}]{Marshall2003}%
  \BibitemOpen
  \bibfield  {author} {\bibinfo {author} {\bibfnamefont {W.}~\bibnamefont
  {Marshall}}, \bibinfo {author} {\bibfnamefont {C.}~\bibnamefont {Simon}},
  \bibinfo {author} {\bibfnamefont {R.}~\bibnamefont {Penrose}}, \ and\
  \bibinfo {author} {\bibfnamefont {D.}~\bibnamefont {Bouwmeester}},\ }\href
  {\doibase 10.1103/PhysRevLett.91.130401} {\bibfield  {journal} {\bibinfo
  {journal} {Phys. Rev. Lett.}\ }\textbf {\bibinfo {volume} {91}},\ \bibinfo
  {pages} {130401} (\bibinfo {year} {2003})}\BibitemShut {NoStop}%
\bibitem [{\citenamefont {Vitali}\ \emph {et~al.}(2007)\citenamefont {Vitali},
  \citenamefont {Gigan}, \citenamefont {Ferreira}, \citenamefont {B\"ohm},
  \citenamefont {Tombesi}, \citenamefont {Guerreiro}, \citenamefont {Vedral},
  \citenamefont {Zeilinger},\ and\ \citenamefont {Aspelmeyer}}]{Vitali2007}%
  \BibitemOpen
  \bibfield  {author} {\bibinfo {author} {\bibfnamefont {D.}~\bibnamefont
  {Vitali}}, \bibinfo {author} {\bibfnamefont {S.}~\bibnamefont {Gigan}},
  \bibinfo {author} {\bibfnamefont {A.}~\bibnamefont {Ferreira}}, \bibinfo
  {author} {\bibfnamefont {H.~R.}\ \bibnamefont {B\"ohm}}, \bibinfo {author}
  {\bibfnamefont {P.}~\bibnamefont {Tombesi}}, \bibinfo {author} {\bibfnamefont
  {A.}~\bibnamefont {Guerreiro}}, \bibinfo {author} {\bibfnamefont
  {V.}~\bibnamefont {Vedral}}, \bibinfo {author} {\bibfnamefont
  {A.}~\bibnamefont {Zeilinger}}, \ and\ \bibinfo {author} {\bibfnamefont
  {M.}~\bibnamefont {Aspelmeyer}},\ }\href {\doibase
  10.1103/PhysRevLett.98.030405} {\bibfield  {journal} {\bibinfo  {journal}
  {Phys. Rev. Lett.}\ }\textbf {\bibinfo {volume} {98}},\ \bibinfo {pages}
  {030405} (\bibinfo {year} {2007})}\BibitemShut {NoStop}%
\bibitem [{\citenamefont {O'Connell}\ \emph {et~al.}(2010)\citenamefont
  {O'Connell}, \citenamefont {Hofheinz}, \citenamefont {Ansmann}, \citenamefont
  {Bialczak}, \citenamefont {Lenander}, \citenamefont {Lucero}, \citenamefont
  {Neeley}, \citenamefont {Sank}, \citenamefont {Wang}, \citenamefont {Weides},
  \citenamefont {Wenner}, \citenamefont {Martinis},\ and\ \citenamefont
  {Cleland}}]{OConnell2010}%
  \BibitemOpen
  \bibfield  {author} {\bibinfo {author} {\bibfnamefont {A.}~\bibnamefont
  {O'Connell}}, \bibinfo {author} {\bibfnamefont {M.}~\bibnamefont {Hofheinz}},
  \bibinfo {author} {\bibfnamefont {M.}~\bibnamefont {Ansmann}}, \bibinfo
  {author} {\bibfnamefont {R.}~\bibnamefont {Bialczak}}, \bibinfo {author}
  {\bibfnamefont {M.}~\bibnamefont {Lenander}}, \bibinfo {author}
  {\bibfnamefont {E.}~\bibnamefont {Lucero}}, \bibinfo {author} {\bibfnamefont
  {M.}~\bibnamefont {Neeley}}, \bibinfo {author} {\bibfnamefont
  {D.}~\bibnamefont {Sank}}, \bibinfo {author} {\bibfnamefont {H.}~\bibnamefont
  {Wang}}, \bibinfo {author} {\bibfnamefont {M.}~\bibnamefont {Weides}},
  \bibinfo {author} {\bibfnamefont {J.}~\bibnamefont {Wenner}}, \bibinfo
  {author} {\bibfnamefont {J.~M.}\ \bibnamefont {Martinis}}, \ and\ \bibinfo
  {author} {\bibfnamefont {A.~N.}\ \bibnamefont {Cleland}},\ }\href@noop {}
  {\bibfield  {journal} {\bibinfo  {journal} {Nature}\ }\textbf {\bibinfo
  {volume} {464}},\ \bibinfo {pages} {697} (\bibinfo {year}
  {2010})}\BibitemShut {NoStop}%
\bibitem [{\citenamefont {Pepper}\ \emph {et~al.}(2012)\citenamefont {Pepper},
  \citenamefont {Ghobadi}, \citenamefont {Jeffrey}, \citenamefont {Simon},\
  and\ \citenamefont {Bouwmeester}}]{Pepper2012}%
  \BibitemOpen
  \bibfield  {author} {\bibinfo {author} {\bibfnamefont {B.}~\bibnamefont
  {Pepper}}, \bibinfo {author} {\bibfnamefont {R.}~\bibnamefont {Ghobadi}},
  \bibinfo {author} {\bibfnamefont {E.}~\bibnamefont {Jeffrey}}, \bibinfo
  {author} {\bibfnamefont {C.}~\bibnamefont {Simon}}, \ and\ \bibinfo {author}
  {\bibfnamefont {D.}~\bibnamefont {Bouwmeester}},\ }\href@noop {} {\bibfield
  {journal} {\bibinfo  {journal} {Physical Review Letters}\ }\textbf {\bibinfo
  {volume} {109}},\ \bibinfo {pages} {023601} (\bibinfo {year}
  {2012})}\BibitemShut {NoStop}%
\bibitem [{\citenamefont {Schumm}\ \emph {et~al.}(2005)\citenamefont {Schumm},
  \citenamefont {Hofferberth}, \citenamefont {Andersson}, \citenamefont
  {Wildermuth}, \citenamefont {Groth}, \citenamefont {Bar-Joseph},
  \citenamefont {Schmiedmayer},\ and\ \citenamefont {Kruger}}]{Schumm2005}%
  \BibitemOpen
  \bibfield  {author} {\bibinfo {author} {\bibfnamefont {T.}~\bibnamefont
  {Schumm}}, \bibinfo {author} {\bibfnamefont {S.}~\bibnamefont {Hofferberth}},
  \bibinfo {author} {\bibfnamefont {L.~M.}\ \bibnamefont {Andersson}}, \bibinfo
  {author} {\bibfnamefont {S.}~\bibnamefont {Wildermuth}}, \bibinfo {author}
  {\bibfnamefont {S.}~\bibnamefont {Groth}}, \bibinfo {author} {\bibfnamefont
  {I.}~\bibnamefont {Bar-Joseph}}, \bibinfo {author} {\bibfnamefont
  {J.}~\bibnamefont {Schmiedmayer}}, \ and\ \bibinfo {author} {\bibfnamefont
  {P.}~\bibnamefont {Kruger}},\ }\href {\doibase 10.1038/nphys125} {\bibfield
  {journal} {\bibinfo  {journal} {Nat. Phys.}\ }\textbf {\bibinfo {volume}
  {1}},\ \bibinfo {pages} {57} (\bibinfo {year} {2005})}\BibitemShut {NoStop}%
\bibitem [{\citenamefont {Clarke}\ and\ \citenamefont
  {Wilhelm}(2008)}]{clarke2008superconducting}%
  \BibitemOpen
  \bibfield  {author} {\bibinfo {author} {\bibfnamefont {J.}~\bibnamefont
  {Clarke}}\ and\ \bibinfo {author} {\bibfnamefont {F.~K.}\ \bibnamefont
  {Wilhelm}},\ }\href@noop {} {\bibfield  {journal} {\bibinfo  {journal}
  {Nature}\ }\textbf {\bibinfo {volume} {453}},\ \bibinfo {pages} {1031}
  (\bibinfo {year} {2008})}\BibitemShut {NoStop}%
\bibitem [{\citenamefont {Zurek}(1981)}]{Zurek1981}%
  \BibitemOpen
  \bibfield  {author} {\bibinfo {author} {\bibfnamefont {W.~H.}\ \bibnamefont
  {Zurek}},\ }\href {\doibase 10.1103/PhysRevD.24.1516} {\bibfield  {journal}
  {\bibinfo  {journal} {Phys. Rev. D}\ }\textbf {\bibinfo {volume} {24}},\
  \bibinfo {pages} {1516} (\bibinfo {year} {1981})}\BibitemShut {NoStop}%
\end{thebibliography}%
\end{document}